\documentclass[conference]{llncs}
\usepackage{graphicx}
\usepackage{caption}
\usepackage{amsmath}
\usepackage{lscape}
\usepackage{enumitem}
\usepackage[lofdepth,lotdepth]{subfig}
\usepackage{comment}
\usepackage{rotating}

\begin{document}

\title{Context-Driven Elicitation of Default Requirements: an Empirical Validation}

\author{Corentin Burnay\inst{1,2,3} \and Ivan J. Jureta\inst{1,2,3} \and St\'ephane Faulkner\inst{2,3}}
\institute{Fonds de la Recherche Scientifique -- FNRS, Brussels \and Department of Business Administration, University of Namur \and PReCISE Research Center, University of Namur \\ \email{$\{$corentin.burnay, ivan.jureta, stephane.faulkner$\}$@unamur.be}}

\maketitle

\begin{abstract}
In Requirements Engineering, requirements elicitation aims the acquisition of information from the stakeholders of a system-to-be. An important task during elicitation is to identify and render explicit the stakeholders' implicit assumptions about the system-to-be and its environment. Purpose of doing so is to identify omissions in, and conflicts between requirements. This paper offers a \textit{conceptual framework} for the identification and documentation of \textit{default requirements} that stakeholders may be using. The framework is relevant for practice, as it forms a check-list for types of questions to use during elicitation. An empirical validation is described, and guidelines for elicitation are drawn.
\end{abstract}


\section{Introduction}
Requirements elicitation is one of the first steps of Requirements Engineering (RE), the main purpose of which is to produce a specification of the system-to-be, which satisfies some requirements, and which is sufficiently clear, precise, and complete to be used in subsequent systems engineering steps. Hereafter, we will refer to requirements elicitation only by elicitation. Elicitation focusses on the acquisition of information from the stakeholders of the system-to-be as a source of requirements that the system-to-be should satisfy. Such process involves communication with the stakeholders. Although communication is not the only means to elicit information relevant for RE, we will focus only on it in this paper.

One important issue during elicitation is that information provided by stakeholders can be uncertain and incomplete. Uncertain, because it reflects their beliefs and desires about the future. Incomplete, because they cannot anticipate all conditions that may arise in the future, when the system-to-be is operational. Our concern in this paper is completeness, while we will not be discussing uncertainty. Regardless of how uncertain information from the stakeholders is, our aim is to look for how to reduce its incompleteness. In particular, our concern in this paper is \textit{how to acquire and document information that is implicitly assumed by stakeholders during elicitation}.

Our starting point is the idea that implicit assumptions that stakeholders make can be understood as a non-monotonic reasoning (NMR). Many influential theoretical frameworks have been proposed to model such reasoning \cite{Mcdermott_1980,Moore_1985,Reiter_1980,Mccarthy_1980}, but we concentrate in this paper on Reiter's default logic, and use it as a conceptualization through which to study implicit assumptions stakeholders may be making during elicitation. We consider that such assumptions are defaults, the normality assumptions that stakeholders consider given, and from which they derive information which they then explicitly communicate. We see defaults as potential source of additional requirements, that we called default requirements.

Our contribution can be summarized in the following points: (i) the review of related work in RE, context and NMR literature, (ii) the presentation of how default logic can be used in RE as a support for the identification of implicit information,(iii) the definition of a context framework to be used as a check-list for documentation during elicitation and (iv) the empirical validation of the framework.

This paper structures as follows. We present a review of related work (\S2), describe the theory of default logic and its application in RE (\S3). We then present a survey of context's definitions and define a framework for the documentation of default requirements (\S4). We describe the way our empirical validation of the framework was handled (\S5), present and discuss the results of the experiment (\S6,7) and provide a final conclusion about our contribution (\S8). 

\section{Related Work}
In their seminal paper on the four dark corners of RE, Zave and Jackson established a core ontology for RE, which describes the important concepts to be accounted for in RE \cite{Zave_1997}. Doing so, they suggest that information about domain assumptions, requirements and specifications of the system-to-be must be collected, documented and analysed in order for RE to be successful. Jureta et al. broaden this ontology by suggesting that any communicated information is relevant to consider as part of requirements problem \cite{Jureta_2009}. In other words, any information that is explicitely adressed to the engineer is relevant to consider . The present paper argues that both explicit \textbf{and} implicit information are relevant to consider.

One common way to identify requirements is the goal-oriented approach, in which engineers should understand the why of a system before defining the what \cite{vanLamsweerde_2001}. Engineers should therefore try to capture intentions of stakeholders for the system-to-be. Various methods exist to capture such information \cite{Coughlan_2002}, some of which focus on the decision-making process of stakeholders \cite{Aurum_2003}. In this paper, we complement such contribution using the more formal default logic theory.

Some attention has been paid to \textit{test empirically the factors that influence default reasoning}. Ford and Billington \cite{Ford_2000} for instance propose an experiment to validate the impact of such subjects-related factors on default reasoning. They present factors such as \textit{the reluctance to draw conclusion based on conflicting rules} or \textit{the number of positive and negative sentences}, which they argue influence the consistence of reasoners. Elio and Pelletier \cite{Elio_1993} propose that default reasoning is likely to be influenced by several external factors. They highlight that whether an objects is \textit{naturally-occurring} or \textit{artificial} influences the way people think about this object. They also discuss the influence of other factors like the \textit{quantity of information} that is provided or the information about the \textit{relative size} of the objects.

To the best of our knowledge, no experiment has been performed in RE to study the use of default reasoning and context during elicitation. Some of our research efforts went on the validation of previous NMR factor in RE \cite{Burnay_2012}, and on the definition of a more complete list of context factors to be accounted for during elicitation \cite{Burnay_2013}. The present paper comes as the conclusion to these initial experiments.

There has been limited attention regarding the question of \textit{accurate context's definition in RE}. Yet, context as a source of information is not new. Many papers propose high level discussions about context in RE: Potts and Hsi \cite{Potts_1997a} emphasize the existence of \textit{Contextualism} -- opposed to abstractionism -- as a possible alternative design philosophy for information systems. Viller and Sommerville \cite{Viller_1999} propose discussions about how \textit{ethnographic analysis} is value-added to RE, thereby broadening the scope of RE context to culture questions. Beyer and Holtzblatt propose the Contextual Design model \cite{Beyer_1998}, which increases the scope of relevant information to any data about \textit{the field where people are living}. Previous works illustrate the trend to include even more data in the scope of RE relevant information. Cohene and Easterbrook \cite{Cohene_2005} discuss a topic closer to what we address in this paper. They suggest that elicitation techniques that are used in an interview should be adapted to fit the kind of information engineers are trying to find, i.e. adapt the elicitation technique to the situation -- or context.

While previous works highlight how valuable information about context is to RE, we find only few papers proposing a structured definition of context that is adapted to RE. One of them is a paper of Sutcliffe et al. \cite{Sutcliffe_2006}, which goes on a method for requirements analysis that aims to accounts for individual, personal goals and the effect of time and context on requirements. They suggest a list of aspects to deal with, but do -- to the best of our knowledge -- no empirical validation. RE community seems to agree on the importance of further research on the link between context and RE. Cheng and Atlee \cite{Cheng_2007} stress the importance of context and empirical validation of RE models as a direction for future research to accelerate the transfer of research results into RE practice.

\section{Default Requirements}
In this section, we introduce a running example, describe what is \textit{default logic} and how it is used to identify implicit information used during requirements elicitation.

\subsection{Running Example: LogisTIC}
LogisTIC is an SME specializing in the distribution of fragile non perishable products (e.g. dishes, decoration, windows, etc.). LogisTIC employs forty people, organized across five departments. One of them concentrates on the management of risk and is responsible, among other, for deciding what transportation company (TC) is the most trustful. For instance, it recently decided to use \textit{Trains \& Co} rather than $Planes\&Co$ because rate of broken parts was significantly larger with the latter. To support the risk management activity, the chief executive officer (CEO) of the firm wishes to accelerate the choice of TCs using a decision-support system (DSS), so that the risk department can focus on other relevant risk aspects. The CEO claims that one of the most critical aspect in this project is to identify what variables must be accounted for in the risk estimation of a TC in order to simulate correctly the decision-making result of a risk employee. A requirements engineer is hired to elicit these variables and specify a system that decides which TC to choose. 

\subsection{What is Default Logic}
NMR theories offer various ways to study decision-making of LogisTIC's risk employees. We consider Reiter's default logic \cite{Reiter_1980} is particularly relevant for this paper, since it builds on concepts that correspond to important aspects of elicitation. Our contribution could however be easily adapted to fit other NMR formalisms such as \cite{Mccarthy_1980,Mcdermott_1980,Moore_1985}. The purpose of Reiter's default logic is to formalize inference rules without explicitly mentioning all their exceptions, i.e. formalize inferences by default of contradictory information. In default logic, the normality assumption of a stakeholder states that, in absence of evidence to the contrary, default assumptions hold. We assume this is what a stakeholder does whenever she has to decide about a requirement facing a context she does not perfectly know: she uses a default theory. 

A default theory is a pair $\left\langle D,W \right\rangle$, in which $D$ is a set of \textit{default rules} $\left\lbrace D_1, ... ,D_n \right\rbrace$, and $W$ is a \textit{background knowledge}. $W$ consists of first order logic (FOL) premises summarizing what the decision-maker knows ``for sure''. $D$ is a set of expressions such as the one reported in Eq. \ref{Default_Definition} that summarizes what the decision-maker believes. Eq. \ref{Default_Definition} reads that if $X$ is a bird, and if it is consistent to assume that $X$ can fly -- $M$ being an operator to evaluate the consistency of the subsequent proposition $Fly(X)$ against the background knowledge $W$ --, then it is believed that $X$ can actually fly. Next sub-section discusses how the previous default theory can be used as a way to model decision-making during elicitation of requirements.

\begin{equation}
D_i = \left\lbrace\frac{Flies(X) : M~Fly(X)}{Fly(X)} \right\rbrace
\label{Default_Definition}
\end{equation}

\subsection{Default Logic in Requirements Engineering}
Back to our running example, we want to support the engineer in determining what aspects are considered by a risk employee when she decides about which TC to choose. We use the default logic to formally document such reasoning.

We denote with $T$ the theory that is used by the employee. $T$ consists of a pair $\left\langle D, W \right\rangle$, and the engineer might be interested in documenting these elements in order to capture requirements and hence specify the DSS. Consider for instance that LogisTIC's employee usually uses \textit{punctuality} as a variable in her decision. This can be documented as an assertion $isOnTime(X)$ -- where $X$ is a TC. Previous assertion can be interpreted by the engineer as the necessity for a TC to be usually on time in order to be selected by LogisTIC's DSS, i.e. as a requirement for the DSS. When she has to decide which TC to choose, the stakeholder reasons about $isOnTime(X)$ using a background knowledge $W$. Assume that she knows the company $Trucks\&Co$ usually suffers from frequent delays, that the company $Trains\&Co$ is usually on time, and that $Planes\&Co$ and $Trains\&Co$ have their own vehicles, which is not the case for $Trucks\&Co$ which shares its vehicles with another company. The previous knowledge background is formalized in Eq. \ref{KStakeholder}.

\begin{equation}
W = \left\{
    \begin{array}{ll}
         ownVhcl(Planes\&Co),~ownVhcl(Trains\&Co),~\neg ownVhcl(Trucks \& Co),\\
         ~\neg isOnTime(Trucks \& Co),~\neg isOnTime(Trains \& Co)
    \end{array}
\right\}
\label{KStakeholder}
\end{equation}

Now, the stakeholder has likely not a perfect knowledge background about the transportation context. There are probably pieces of information that she does not know, which are required for her to make a decision about which TC to choose. For instance, Eq. \ref{KStakeholder} tells us that the stakeholder does not know whether $Planes\&Co$ is usually on time. Because this information is still necessary to solve $isOnTime(Planes\&Co)$, the stakeholder may fill in this gap by assuming that TCs having vehicles of their own are usually on time, unless such assertion is not consistent with what she knows. In other words, the stakeholder can use $W$ in combination with a set of default rules $D$, and can thereby build a default theory $T$. The assertion $isOnTime(Planes\&Co)$ consequently suggests the use of a default rule such as reported in Eq. \ref{DStakeholder}.

\begin{equation}
D = \left\lbrace\frac{ownVhcl(Planes\&Co) : M~isOnTime(Planes\&Co)}{isOnTime(Planes\&Co)} \right\rbrace
\label{DStakeholder}
\end{equation}

$D$ reads as follows: if $Planes\&Co$ has its own vehicles, and if it is consistent to believe that $Planes\&Co$ is usually on time, then it can be believed that $Planes\&Co$ is usually on time. The member $ownVhcl(Planes\&Co)$ is a \textbf{prerequisite} to the default: it is necessary but not sufficient for the default to hold. The member $:M isOnTime(Planes\&Co)$ is the \textbf{consistency test} that must also be verified for the rule to hold. As a reminder, $isOnTime(Planes\&Co)$ is consistent with $W$ if $\neg isOnTime(Planes\&Co)$ does not follow from $W$ based on deductive inferences. The denominator member $isOnTime(Planes\&Co)$ is the result of a default rule, and is a \textbf{belief} of the stakeholder.

The way we see previous situation is that a default theory suggests larger requirements than what is communicated by a stakeholder. For LogisTIC, the employee reasons about the initial assertion $isOnTime(X)$ using an incomplete $W$. Doing so, she defines a new default rule, the belief of which being her initial assertion, e.g. based on Eq. \ref{DStakeholder}, she decides it is reasonable to believe that $Planes\&Co$ is usually on time because it has vehicles of its own. Using such rule, the stakeholder suggests that it is important for the DSS to account in some way for the fact that a TC has its own vehicles. In the case such default rule is not documented, the engineer would work with a single requirement $isOnTime(X)$ and specify a system that does not behave exactly as the risk employee reasons. We interpret the previous as the existence of a \textit{default requirement}.

\subsection{What are Ground and Default Requirements}
A stakeholder may communicate an assertion for which she has a perfect knowledge. In such case, reasoning about that assertion does not imply to use a default theory. \textbf{Ground requirement} refers to such assertion, for which it is not necessary to use a default rule in order to reason about. A ground requirement is an assertion on which the stakeholder can decide strictly based on $W$. It takes the form of Eq. \ref{GroundRequirement}.

\begin{equation}
ground~requirement = assertion(X)
\label{GroundRequirement}
\end{equation}

We call \textbf{default requirement} any default rule that is used when the stakeholder has imperfect knowledge background to reason about the assertion. Using a default requirement, the stakeholder communicates a belief -- previously referred to as \textit{assertion} and assumes some underlying ground requirements -- previously referred to as \textit{prerequisites}. We adopt the view that a belief is a requirement for the system when it is communicated by means of directive speech acts \cite{Jureta_2009}. Both belief and ground requirements form source of information about expectations of the stakeholder toward the system-to-be. The requirement is said to be ``by default'' because it describes an assertion that is believed as long as its ground requirements are verified. A default requirement takes the form of Eq. \ref{DefaultRequirement}. $X$ is called an \textit{object}, and can be seen as the entity targeted by the default requirement. 

\begin{equation}
default~requirement(X) = \left\lbrace\frac{ground~requirements(X) : consistency~test(X)}{belief(X)} \right\rbrace
\label{DefaultRequirement}
\end{equation}

Based on equation \ref{DefaultRequirement}, an engineer knows that any belief communicated via directive speech act is not complete enough to be reported as a single requirement. To improve completeness of the elicitation, engineers might want to identify the complete default rule that has been used by the stakeholder in order to decide about the communicated belief: hence, the engineer could identify underlying ground requirements. We believe the previous supports engineers in that the use of a default rule by a stakeholder may not always be explicitly communicated, thereby preventing the engineer from identifying the underlying ground requirements of a belief. We see many reasons for this, the most significant ones being that:

\begin{itemize}
	 \item stakeholders may consider the rule is not relevant to the engineer;
	 \item stakeholders may be reluctant to share their rules;
	 \item stakeholders may not be aware they use default rules.
\end{itemize}

\subsection{Cascading Default Requirements}
Most of the time, default requirements will be judged as relevant by a stakeholder, who will therefore communicate spontaneously some beliefs and their related ``ground requirements''. What is important to consider is that doing so, the stakeholder turns default requirements -- and more precisely the ground requirements of the default requirement -- into new additional and finer assertions. As previously explained, these new assertions may be themselves default or ground requirements, depending on the stakeholder's knowledge background. There is therefore a cascading effect, in the sense that a default requirements builds on assertions that may themselves form default or ground requirements. The responsibility of an engineer is then to go sufficiently downward in the default tree to identify the requirements for which the stakeholder has a complete background and does not need to use a default rule, i.e. to identify ground requirements.

Considering the case of LogisTIC, it is likely that the stakeholder will share her default requirement $D$ according to which $isOnTime(X)$ can be decided based on $ownVhcl(X)$: it is an important characteristics that she won't omit to share. Doing so, the stakeholder introduces a new assertion $ownVhcl(X)$. The question that must then be answered is whether the stakeholder has still imperfect knowledge about that new assertion -- in which case she will likely use a new default requirement -- or whether she can decide about that requirement based on her knowledge -- in which case the engineer has achieve the actual ground requirement. Assuming the first case, the stakeholder could then adopt a new rule according to which a TC has its own vehicles as long as it is relevant to believe so, and if this TC has reported the vehicles in its balanced sheet. For an engineer, this would result in a new default rule $D'$ to document, such as reported in Eq. \ref{DStakeholderBis}. Assuming the second case, the engineer can assume the elicitation is complete enough.

\begin{equation}
D' = \left\lbrace\frac{\left\lbrace\frac{vhclInBalancedSheet(X) : M~ownVhcl(X)}{ownVhcl(X)} \right\rbrace : M~isOnTime(X)}{isOnTime(X)} \right\rbrace
\label{DStakeholderBis}
\end{equation}

It is worthwhile to note that such cascading process -- which generates default requirements -- is different from classical requirements decomposition which generate sub-requirements \cite{Yu_1997}. Default and sub-requirements differ in that: 

\begin{itemize}
	\item Defaults are de-feasible;
	\item Defaults are beliefs, suppositions or assumptions related to a requirement;
	\item Defaults are transversal, not hierarchically related to a requirement.
\end{itemize}

\subsection{Documenting Default Requirements}
The elicitation of default requirements can be achieved through the systematic discussion of assertions communicated by a stakeholder. That is, an engineer should not limit the elicitation process to a discussion of what the stakeholder says, but should also discuss about what potential default rules have been used that are at the origin of a communicated assertion. Such elicitation can be performed using well known requirements elicitation techniques such as those suggested in \cite{Goguen_1993}.

Elicitation is however not complete without an adequate documentation of collected ground and default requirements. Such documentation for ground requirements is straightforward, as it takes the form of a single assertion. Documentation of default requirements likely raises more important conceptual questions, mainly because default requirements are rules, not assertions. One particularly relevant concern is the one of validity: does a default always form a \textit{valid} source of information about requirements of a stakeholder? Are there particular conditions in which the default applies? What would happen if these conditions are not respected? 

In the example of LogisTIC, we identified a default requirement in Eq. \ref{DStakeholder}, which was suggesting that $ownVhcl(X)$ must be accounted in some way when deciding about $isOnTime(X)$. Yet, the default rule was initially assumed for an object $X=Planes\&Co$. In fact, it may not have been the intention of the stakeholder to believe Eq. \ref{DStakeholder} for $X=Trains\&Co$. The question that must then be answered is to know what are the conditions that must be respected in order for the default requirement to be applicable. For instance, the default may have been designed to apply only to TCs using flying vehicles, thereby invalidating the default as a source of requirement about $Trains\&Co$. We consider that, given the nature of default rules, it is relevant to describe in the documentation what are the conditions that must be verified in order for the default rule to be valid, i.e. define the \textit{default domain}. 

Let $X^D$ refer to such default domain, for the default $D$. One possible way of documenting $X^D$ is to list all the possible values of $X$ that respect previously discussed conditions. For LogisTIC, it would result in the list of all TCs using flying vehicles:

\begin{equation}
X^D=\lbrace Planes\&Co,Zeppelins\&Co,Helicopters\&Co,...\rbrace
\label{DDomain}
\end{equation}

As such list can be virtually infinite, this solution seems to be infeasible. An alternative solution consists in defining a set of conditions that were applying at the moment the default requirement was used, and assuming that the default requirement is a valid source of requirements about objects as long as the application of the rule for that object respects these conditions. Doing so, an engineer would ultimately end up with a list of ground requirements to be completed with a list of default requirements and the conditions under which they apply. An example of formal documentation for the requirement of LogisTIC's risk employee about the DSS is reported in Table \ref{DocumentationDSS}. Note that the documentation of conditions is not required for ground requirements as they are objective propositions. As such, they imply no reasoning and are expected to be interpreted by other stakeholders in a same consistent way. This is in sharp contrast with default requirements, which are inherently referring to subjective reasoning, and may not be understood in the same way by other stakeholders of the system. Next section defines what conditions there are that should to be accounted for when documenting default requirements.

\begin{table}[!t]
\begin{center}
\begin{tabular}{|c|p{7cm}|}
\hline 
Proposition & Description \\
\hline
 $isOnTime(X)$
 & \textit{Ground Requirement} where $X$ = any TC \\ \hline

 $\left\lbrace\frac{ownVhcl(X) : M~isOnTime(X)}{isOnTime(X)} \right\rbrace$
 & \textit{Default Requirement} where $X$ = any TC which respects conditions stated in $X^D$\\ 
\hline 
\end{tabular} 
\caption{Documentation of Requirements for Default Rule $D$ (Eq. \ref{DStakeholder}) \label{DocumentationDSS}}
\end{center}
\end{table}

\section{Context in Requirements Engineering}
The validity of a default rule is ensured only for objects that respect conditions described in the so-called default domain. This domain restricts the validity of a default requirements to objects respecting the conditions of that requirement. This section argues that conditions of a default requirement can be completely characterized by the \textit{context} in which the default was initially formulated. It reviews existing definitions of context and suggests a \textit{classification of conditions} that should be accounted for when documenting default domains.

\subsection{Survey of Context Definitions}
Given our objective to achieve more complete elicitation and documentation, we want to account for the largest possible set of default conditions: this implies to account for the object on which the rule applies, but also for any information related in some way to that object. In this paper, we consider context is an adequate way to document conditions. Context refers to any relevant information that is related in some way to an object. The purpose of this section is to review definitions of context from ubiquitous / context-aware computing literature, and to identify salient dimensions of context that, we argue, also suggest some important conditions to be documented in default domain.

For the sake of usability in RE, we want the conditions to be categorized into dimensions. The definitions we treat in this section are therefore operational: they provide a set of dimensions that are assumed to operationalize context when instantiated, in contrast to lexical definitions that simply describe from a theoretical point of view what context consists of. Table \ref{tab1} summarizes these definitions and presents the dimensions suggested by the latter (X refers to dimensions that are explicitly treated in the article, ? refers to categories that are suggested, but not explicitly mentioned).

\begin{table} [!t]
	\centering
		\begin{tabular} {lccccccccc} \hline \noalign{\smallskip}
		 & \begin{sideways} Place and Time \end{sideways} & \begin{sideways} Individuals \end{sideways} & \begin{sideways} Resources \end{sideways} & \begin{sideways} Physical Conditions \end{sideways} & \begin{sideways} Knowledge \end{sideways} & \begin{sideways} Relationship \end{sideways} & \begin{sideways} Activity and Goal \end{sideways} & \begin{sideways} Computer State \end{sideways} & \begin{sideways} Imaginary Companions \end{sideways}\\ \noalign{\smallskip} \hline \noalign{\smallskip}
		Schillit \& Theimer \cite{Schilit_1994b} & X & X & X &   &   &   &   &   &   \\
		Schilit et al. \cite{Schilit_1994a}  & X & X & X & X &   & ? &   &   &   \\
		Brown \cite{Brown_1996}                 & X & X & ? & X &   & ? &   & X & X \\
		Abowd et al. \cite{Abowd_1997}          & X & X & ? &   & ? &   &   &   &   \\
		Lenat \cite{Lenat_1998}                 & X & X & X &   & X &   &   &   &   \\
		Dey et al. \cite{Dey_2001b}          & X & X & X & X &   &   &   &   &   \\
		Dey \cite{Dey_2001a}                     & X & X & X &   &   & X & ? &   &   \\
		Zimmermann et al. \cite{Zimmermann_2007}    & X & X & X &   &   & X & X &   &   \\
		 \noalign{\smallskip} \hline \noalign{\smallskip}
		\end{tabular}
\caption{Dimensions of Context}
	\label{tab1}
\end{table}

We observe that all of the definitions concur in their observation that \textit{time} and \textit{space} are dimensions of the operational definition of context, characterizing the localization of a context on a time-line / map. The same agreement is observed for the \textit{individual} dimension, which deals with people in the context. Nearby \textit{resources} is another dimension that is regularly defined as part of the context, and which deals with animals and non-living things such as materials, objects, or any other artefacts that can be accessed by individuals and with which they may interact. 

While the large majority of definitions deals with ``things'' and their localization, it is much more difficult to find an agreement on the remaining dimensions that form context. The \textit{physical conditions} are sometimes considered as relevant for the definition of context. \textit{Knowledge} is a less common dimension which includes elements about the content of knowledge, its justification and how it is evaluated by individuals. The notion of \textit{relationship} between individuals is another recurrent element in literature on context, and deals with the relations that exist between two or more individuals or resources. \textit{Activity} is a dimension dealing with the goals / intentions of individuals: what do they aim to do in the context? \textit{Computer's state} and \textit{Imaginary companions} are other plausible dimensions of context.

\subsection{Dimensions of Context}
\begin{figure} [!ht]
\begin{center}
\includegraphics [scale=0.26]{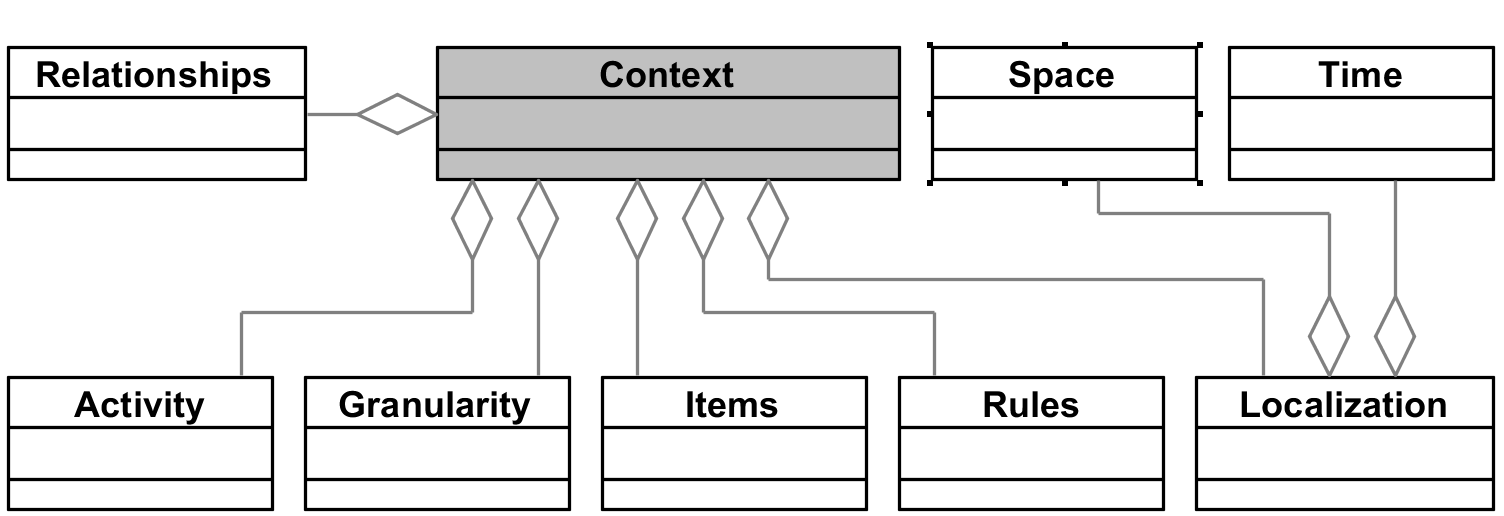}
\caption{Context Framework}
\label{Context_Framework}
\end{center}
\end{figure}

Based on our survey, it is possible to build a framework that supports engineers during elicitation of default requirements. The framework lists dimensions of context that can be used to describe a default domain, and is therefore to be used as a check-list against which to study a default requirement. Among other, the framework emphasizes what conditions have not been considered by an engineer and which should be documented. The framework takes the form of a taxonomy, inspired by our literature review, and adapted to fit the specifics of requirements elicitation and default domains. It includes six important dimensions: (i) Items, (ii) Rules; (iii) Localization, (iv) Activity, (v) Relationship and (vi) Granularity. Based on these six dimensions, it is possible to account for characteristics suggested by each of the reviewed context dimension. The framework is represented in Figure \ref{Context_Framework}. As a way to consolidate the framework, ensure no important dimension has been omitted and to support the interpretation of the framework, we mapped the dimensions with Bunge's ontology on ``the furniture of the world'' \cite{Bunge_1977} (Table \ref{Bunge}).

\begin{table} [!ht]
\begin{center}
\begin{tabular}{|p{3.3cm}|p{4cm}|p{2.2cm}|}
\hline 
\multicolumn{2}{|c|}{\textbf{Bunge's Ontology}} & \textbf{Framework} \\ \hline
\textit{Concept} & \textit{Description} & \textit{Dimension} \\ \hline
Things & fundamental concept & Items  \\ \hline

Properties (intrinsic, mutual, ...), States (functions, schema) & things have properties that describe them. States describe properties & Activities Granularities Localization \\ \hline

Laws & Restriction on or relation of properties  & Rules  \\ \hline

Compositions & Things can be composed to form composite things & Relationships  \\ 
\hline
\end{tabular}
\caption{Mapping our Context Framework with Bunge's Ontology}\label{Bunge}
\end{center}
\end{table}

To illustrate the documentation of validity conditions for default requirements, we use our introductory example. As a reminder, an engineer has been hired to elicit requirements about a DSS that will support the company in deciding in an more automated way which TC to use for the transportation of its products. During elicitation, the engineer discovered that the employee was using a default rule, according to which the punctuality of $Planes\&Co$ can be decided based on whether it has vehicles of its own. Based on this observation, the engineer documented a default requirement $D$, and tries to understand under which conditions such default requirement actually holds true. The conditions identified using the six dimensions of our framework are summarized in an improved version of the initial documentation and illustrated in Table\ref{DocumentationDSSBis}.

\subsubsection{Items}
Items deal with characteristics related to salient entities existing inside the context. It suggests mainly questions about intrinsic properties and behaviour of the object.

\begin{example}\label{ex:item}
Knowing about Items, the engineer adapts her questions to discover that the default requirement only applies to TCs which use flying vehicles, e.g. $useFlyingVhcl(X)$. This condition restricts the use of $D$ to TCs using flying vehicles because the stakeholder believes other TCs are more likely to be stuck in traffic jam and hence face punctuality issues.
\end{example}

\subsubsection{Rules}
Rules deals with constraints that exists in the context and which influence in some way the actions of an item. It suggests mainly questions about laws, cultures, habits, environmental conditions etc. influence the behaviour of the object.

\begin{example} \label{ex:Rule}
Knowing about the Rules dimension, the engineer adapts her questions to discover that $D$ only applies for the objects that respects the quality charter developed by LogisTIC, e.g. $qualityChart(X)$. This condition restricts the use of $D$ to quality TCs because the stakeholder believes TCs which have no lean processes have a more important risk of being late.
\end{example}

\subsubsection{Localization} 
Localization deals with the position of an object. Localization divides into two subcategories: one relating to the time when the object occurs, the other dealing with place where the object occurs.

\begin{example} \label{ex:loc}
Knowing about Localization, the engineer discovers that $D$ only applies for the TC that exist nowadays in Europe, e.g. $locatedInEurope(X)$. This condition suggests that $D$ cannot be used to estimate punctuality of a TC located in Asia because the stakeholder believes Asian TCs may have different priorities.
\end{example}

\subsubsection{Activity} 
Activities deal with the set of objectives of Items. It suggests mainly questions about the goals, the desires and the intentions of Items.

\begin{example}
Knowing about Activities, the engineer discovers that $D$ only applies for the TCs focusing on the transportation of material, e.g. $freightTransport(X)$. This condition restricts the use of $D$ to freight TCs because the stakeholder believes other TCs have different processes resulting in longer delay.
\end{example}

\subsubsection{Relationship} 
Relationships deal with the connections / links between Items and/or Rules. They focus on the way objects relate to each other. It suggest mainly questions about the type, the nature, the direction and the strength of the relations.

\begin{example}
Knowing about Relationships, the engineer discovers that $D$ only applies for the TCs that have no collaboration with other TCs, e.g. $\neg collaborates(X)$. This condition restricts the use of $D$ to competing TCs because the stakeholder believes collaboration may imply unexpected re-allocation of resource that could lead to delays.
\end{example}

\subsubsection{Granularity} 
Granularity deals with the nature, the quantity and the level of any additional piece of information that is provided about objects.

\begin{example}
Knowing about Granularity, the engineer discovers that $D$ only applies for the TCs having vehicles that on average are not older than three years, e.g. $hasNewEquipment(X)$. This condition restricts the use of $D$ to TCs having modern material, because the stakeholder believes vehicles older than three years present a more significant risk of breakdown, and therefore of delay.
\end{example}

\begin{table}[!ht]
\begin{center}
\begin{tabular}{|c|p{7cm}|}
\hline 
Proposition & Description \\
\hline
 $isOnTime(X)$
 & \textit{Ground Requirement} where $X$ = any TC \\ \hline

 $\left\lbrace\frac{ownVhcl(X) : M~isOnTime(X)}{isOnTime(X)} \right\rbrace$
 & \textit{Default Requirement} where $X$ is a TC s.t.:
$hasFlyingVhcl(X) \wedge
qualityChart(X)\wedge
locatedInEurope(X)\wedge
freightTransp(X)\wedge
\neg collaborates(X)\wedge
hasNewEquipment(X)$ \\
\hline 
\end{tabular} 
\caption{Improved Documentation of Requirements for Default Rule $D$ (Eq. \ref{DStakeholder}) \label{DocumentationDSSBis}}
\end{center}
\end{table}

\section{Empirical Validation of Objects Characteristics}
This section describes the empirical study that has been performed in order to validate the context framework as a guideline for the identification of conditions that must be respected in order for a default requirement to hold true. 

\subsection{Design}
The validation of the framework relies on answers of decision-makers to six distinct questionnaires, each dealing with one dimension of the framework. Questionnaires test whether variations in conditions actually alter the outcome, and therefore the validity, of a default requirement. Questionnaires consist of three parts: a \textit{context statement}, a \textit{assignment} and a \textit{variability area}.

\subsubsection{Context Statement}
Subjects are first asked to carefully read the context statement. The statement describes a basic situation to subjects, close to the one we discussed in our LogisTIC running example. The statement describes a regular benchmark problem that requires \textit{basic default reasoning} to be used \cite{Lifschitz_1989}. A benchmark is a problem in which at least two objects are introduced, that are supposed to respect a rule. Subjects are then informed that one object does actually not respect the rule, i.e., there is an exception to the rule. Based on this statement, subjects are asked whether the remaining object respects that rule. The referent context used in our experiment is:

\begin{quotation}
``A survey established by a Belgian national daily newspaper, is published on January 31, 2009. The survey proposes a comparison of some logistic means. Among others, it compares the supply of goods by rail or by boat. The newspaper asserts both are means of transport with transport areas of their own. The newspaper also asserts that types of transport that have their own infrastructures (waterways, railways, etc.) usually offer a reliable service: little delay, no risk of damage, etc. Finally, the newspaper emphasizes that these types of transport were originally created for logistic, which has led to an incredibly strong competition between them.

One year later, in 2010, the newspaper puts forward that for some time, deliveries by train arrived regularly with a delay and more significant losses. The newspaper has not had the opportunity to repeat a study about the boats. As a risk employee in a logistics company, you must decide whether you can trust boat transport.''
\end{quotation}

The referent context can be summarized with the following default benchmark problem: logistic means with transport areas of their own provide typically good service quality (\textit{\footnotesize default rule}). Trains (\textit{\footnotesize object 1}) and boats (\textit{\footnotesize object 2}) have transport areas of their own. Trains provide poor service quality (\textit{\footnotesize exception}). What quality is provided by boats? (\textit{\footnotesize benchmark}). The answer that is expected is that boats provide quality service, i.e. that the exception has no bearing on the default rule. The previous problem can be solved using a default rule similar to the one introduced in our running example, such as:

\begin{equation}
D = \left\lbrace\frac{ownInf(boats) : M~isQuality(boats)}{isQuality(boats)} \right\rbrace
\label{DExperiment}
\end{equation}

\subsubsection{Answers}
Using the referent context, subjects are asked to make a decision about quality service of boats. Since the referent context is built according to the benchmark problem structure, the choice offered to subjects is limited to four different options. For each element inside the variability area, subjects are asked to select one answer.

\begin{itemize}
\item Benchmark: the exception has no bearing on the remaining object;
\item Exception: the exception also applies to the remaining object;
\item Other: the exception imply another exception, with different characteristics;
\item Can't Say: the subject cannot choose one of the former propositions.
\end{itemize}

\subsubsection{Variability Area}
The variability elements are introduced in the last part of the questionnaire. Subjects are asked to answer to the question ``what about the trains?'', considering turn by turn each of the $n$ values of the dimension we want to test. Element $1$ is the value that is presented in the referent context. Element $i (i= 1 ... n-1)$ are substitute values that modify the referent context. Table \ref{variability} presents these elements for each dimension of the framework (the letter R refers to the referent value).

The $n$ parameter has been carefully considered since a trade-off exists:  little n reduces the possibility to interpret the influence of category changes; large n makes the questionnaires too long, with a risk a bias/lack of motivation of subjects. In our experiment, $n$ is set to 4. It is worthy to note that a same questionnaire always alters at most one dimension. Thereby, if an impact on the application of the default is observed, it can only be explained by the varying value of the dimension. 

\begin{table} [ht!]
\begin{center}
\begin{tabular}{|p{1.78cm}|p{9.6cm}|p{0.08cm}|}
  \hline \textbf{Dimension} & \textbf{Variability Area} &  \\
  \hline
  Items & 1. X is a train and Y is a boat & R \\
  54 subjects   & 2. X is a new teleportation technology and Y is a plane &\\
        		& 3. X is a helicopter and Y is a plane &\\      
 				& 4. X is a sub-contractor and Y is the national postal service&\\ \hline
 				  Rules& 1. X and Y are evaluated by a national daily newspaper &R \\
  52 subjects   & 2. X and Y are evaluated by a national management journal &\\
        		& 3. X and Y are evaluated by international management journal &\\      
 				& 4. X and Y are evaluated by a local people magazine&\\ \hline
 				
 				  Localization& 1. X and Y exist in Belgium, nowadays &R \\
  45 subjects   & 2. X and Y exist in China, nowadays &\\
        		& 3. X and Y existed in 1948 &\\      
 				& 4. X and Y will exist in 2025 &\\ \hline
 				
 				  Activity & 1. X and Y were originally created for freight &R \\
  45 subjects   & 2. X and Y were originally created to transport persons &\\
            	& 3. X and Y were originally created to transport any kind of load &\\ 
 				& 4. X and Y were originally created for leisure &\\ \hline
 				
 				  Relation & 1. X and Y strongly compete with each others &R \\
  32 subjects   & 2. X and Y compete normally with each others &\\
 				& 3. X and Y strongly collaborate with each others &\\
 				& 4. X and Y keep neutral relationships &\\  \hline
 				
 				 Granularity & 1. X has one hour late with important damages &R \\
  33 subjects   & 2. X has one day delay and huge losses &\\
          		& 3. X and Y are very similar in their way of working &\\      
 				& 4. X and Y are very different in their way of working &\\ \hline	
\end{tabular}
\caption{Tested Dimensions Levels by Dimension} \label{variability}
\end{center}
\end{table}

\subsection{Subjects}
A total of 260 subjects were questioned. All subjects were students in management sciences, economics or computer sciences from the University of Namur. Each subject answered four different questions, bringing the total number of treated data to approximatively 1000 decisions. Details about respective size of samples are presented in the results section. Each group had twenty minutes to read the referent context definition and answer the questions. Subjects were not compensated for participating in the study, and were asked to answer during class time.

\subsection{Procedure and Validity}
Subjects can use any kind of material. Assignment mentions there is no best answer and tells subjects that the objective of the questionnaire is to understand how managers reason in a situation of imperfect information. 

Validity concerns have been carefully study. We paid attention to eliminate threats to (i) \textbf{external validity} -- addressing the ability to generalize results of the experiment to other people and other situations--, (ii) \textbf{construct validity} -- ensuring that the experiment measures what must actually be measured -- (iii) \textbf{internal validity} -- dealing with the extent to which the independent variables causally link to the dependent variables -- and finally (iv) \textbf{conclusion validity} -- aiming to ensure that there is a significant relationship between the treatment and the outcome of the survey \cite{Wohlin_2012}. We consider the use of students is not a threat to validity, as our main concern is to test human reasoning.

\newpage
\section{Results}
\begin{figure}
\centering
\vspace{-10mm}
\subfloat{\includegraphics [trim= 2cm 1.0cm 3cm 0cm,clip, scale=0.20]{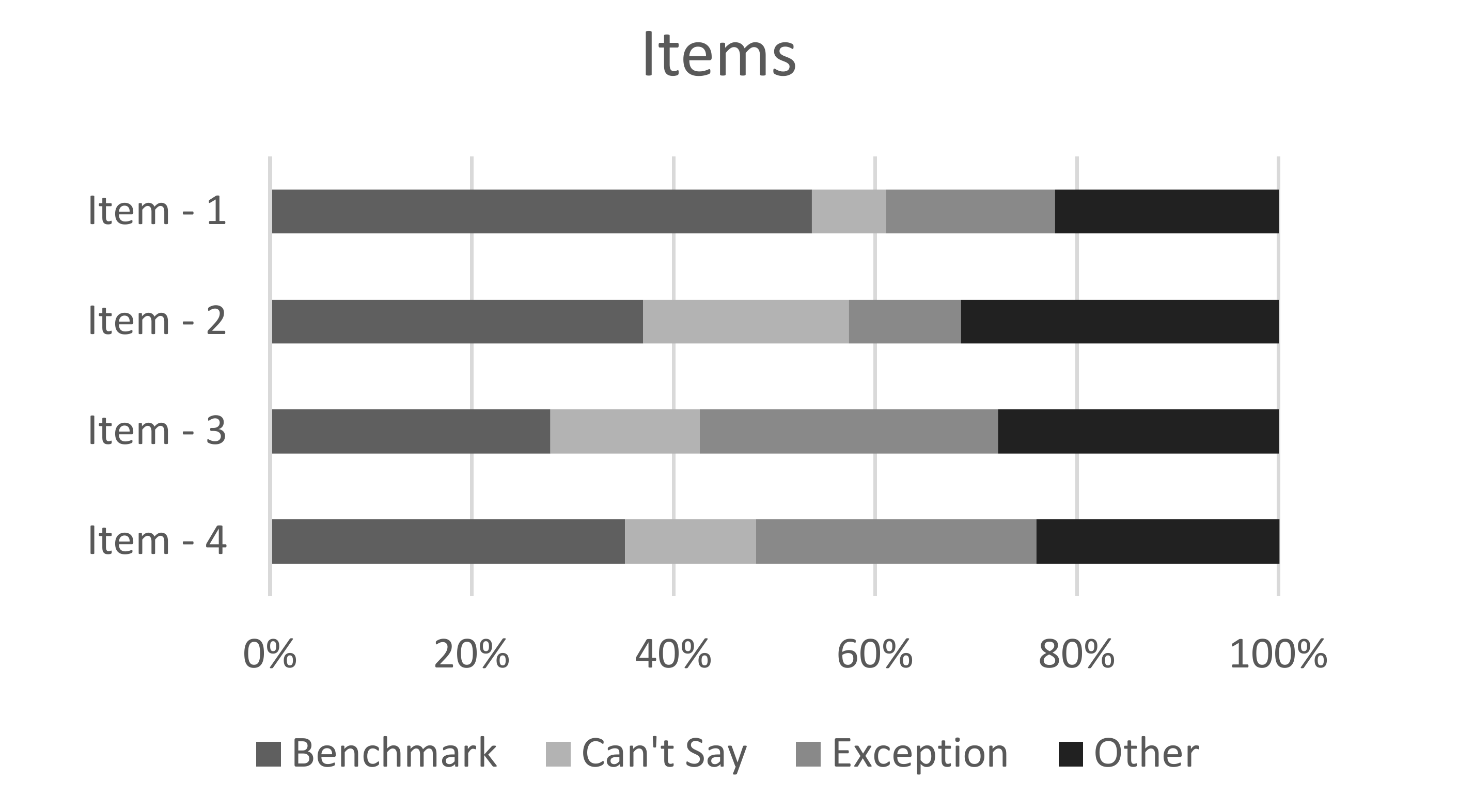}}
\qquad
\subfloat{\includegraphics [trim= 2cm 0.8cm 3cm 0cm,clip,scale=0.20]{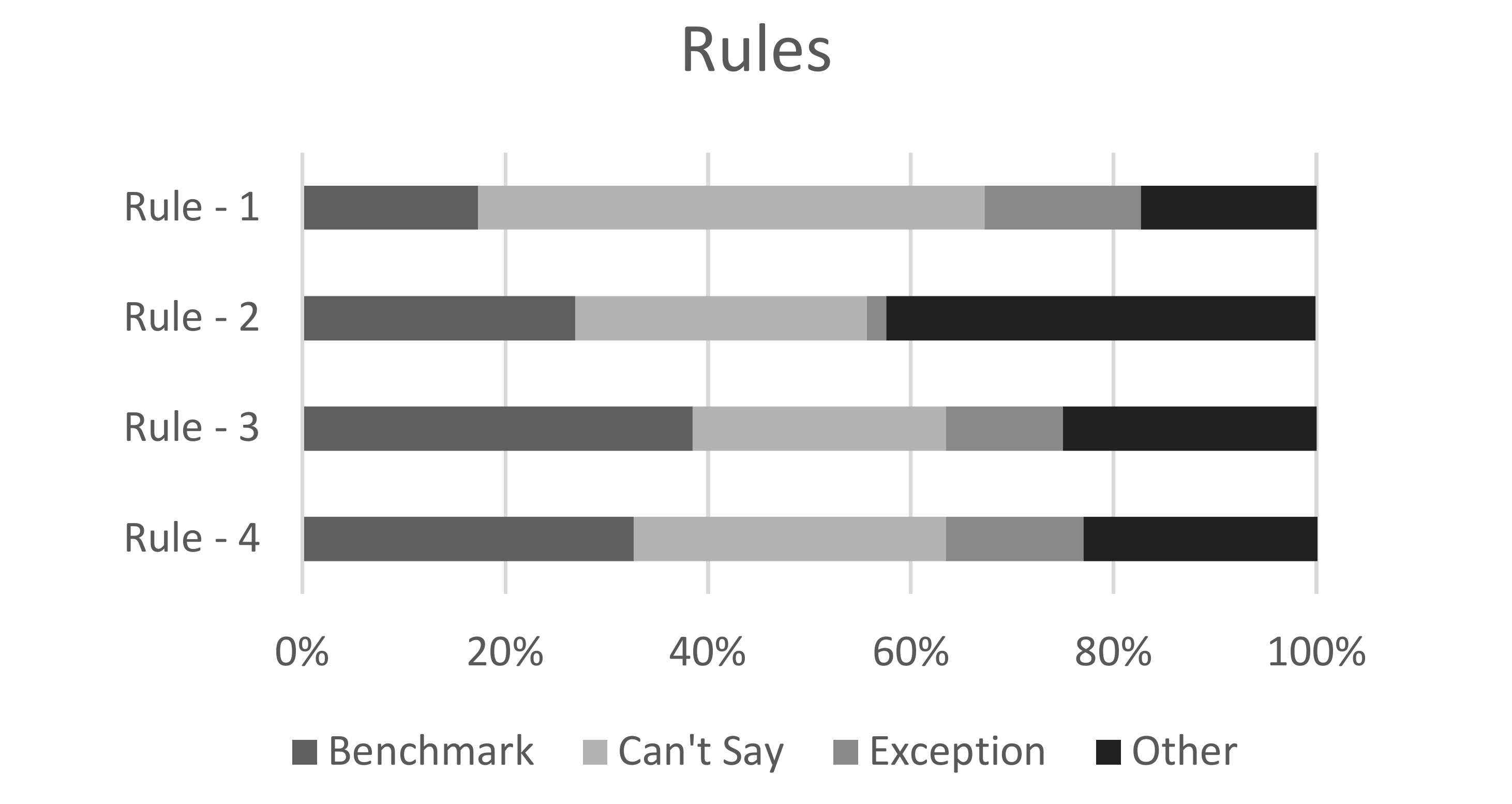}}    
\qquad
\subfloat{\includegraphics [trim= 1.8cm 0cm 3cm 0cm,clip,scale=0.20]{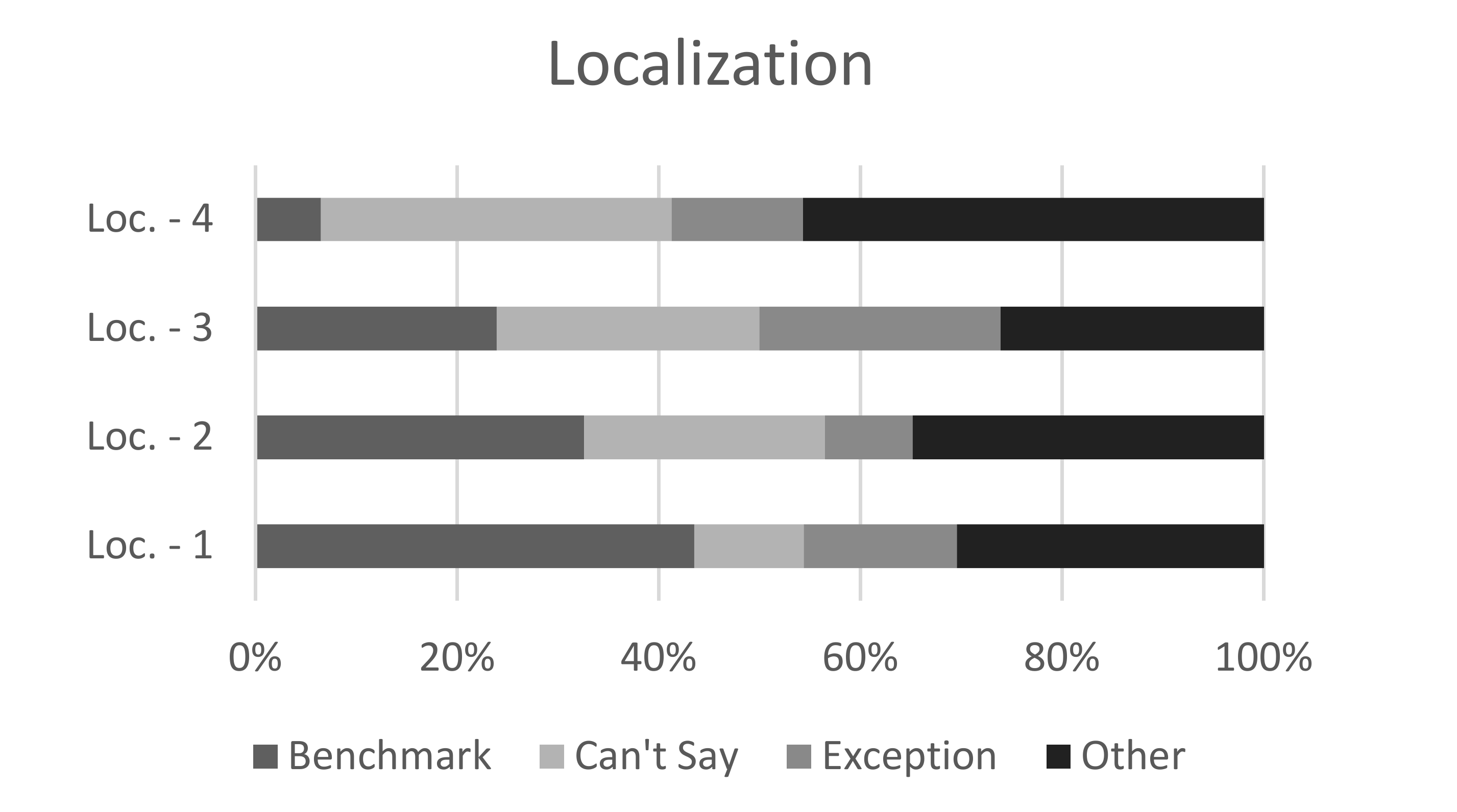}}       
\qquad
\subfloat{
		\includegraphics [trim= 2.5cm 0cm 3cm 2.0cm,clip,scale=0.20]{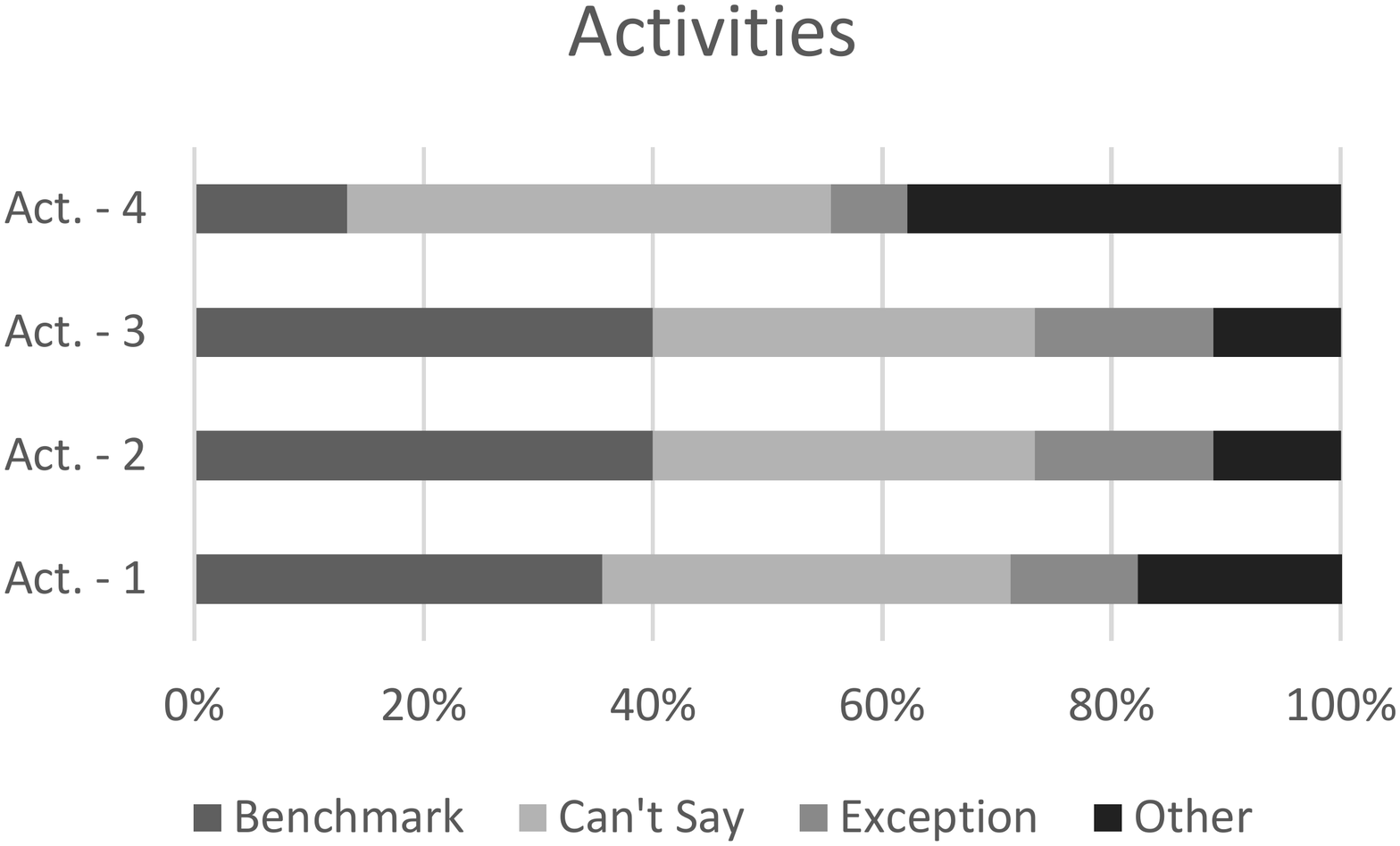}}
\qquad
\subfloat{\includegraphics [trim= 2cm 0cm 3cm 0cm,clip,scale=0.2]{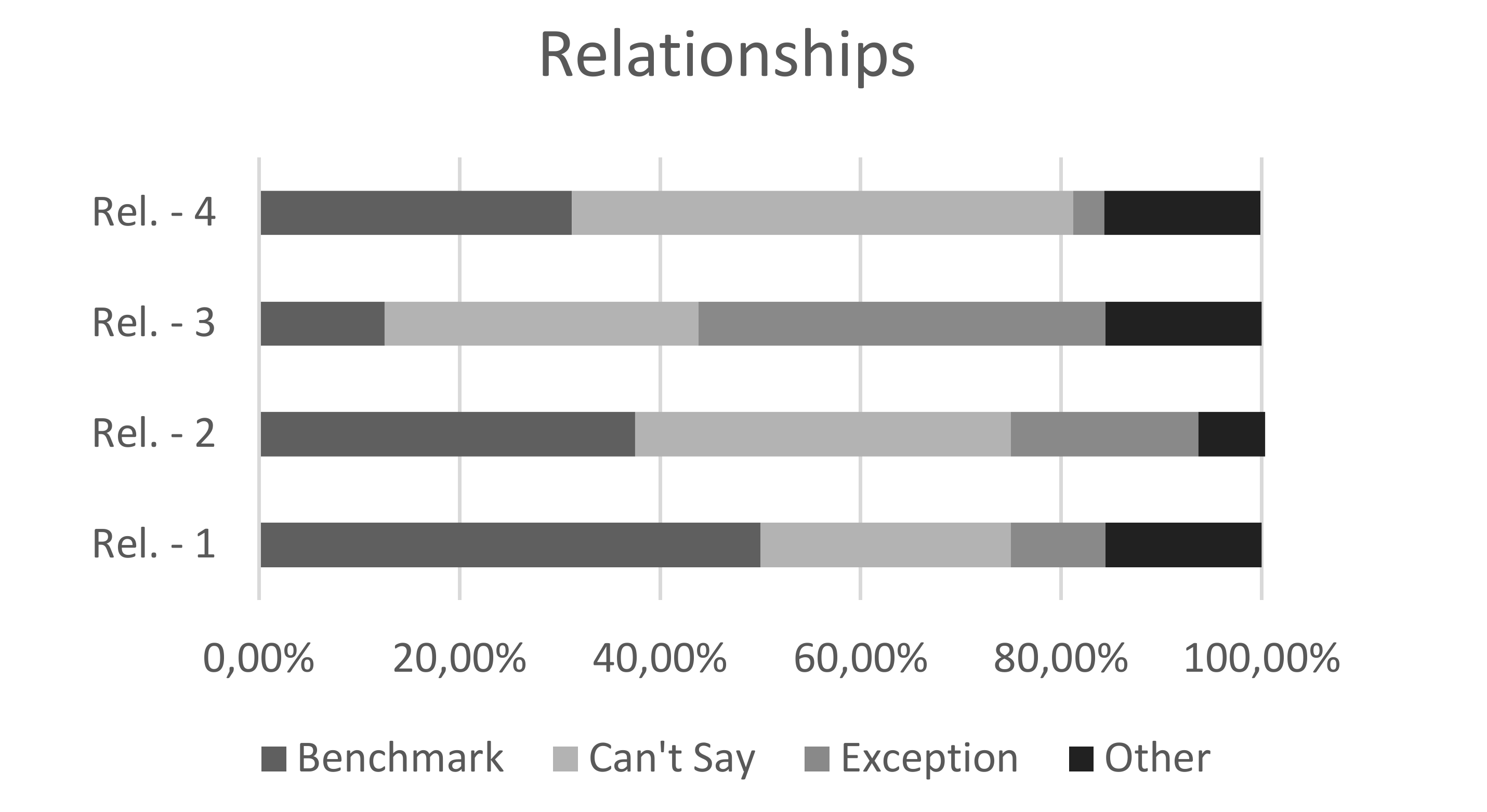}}
\qquad
\subfloat{
          \includegraphics [trim= 2cm 0cm 3cm 0cm,clip,scale=0.2]{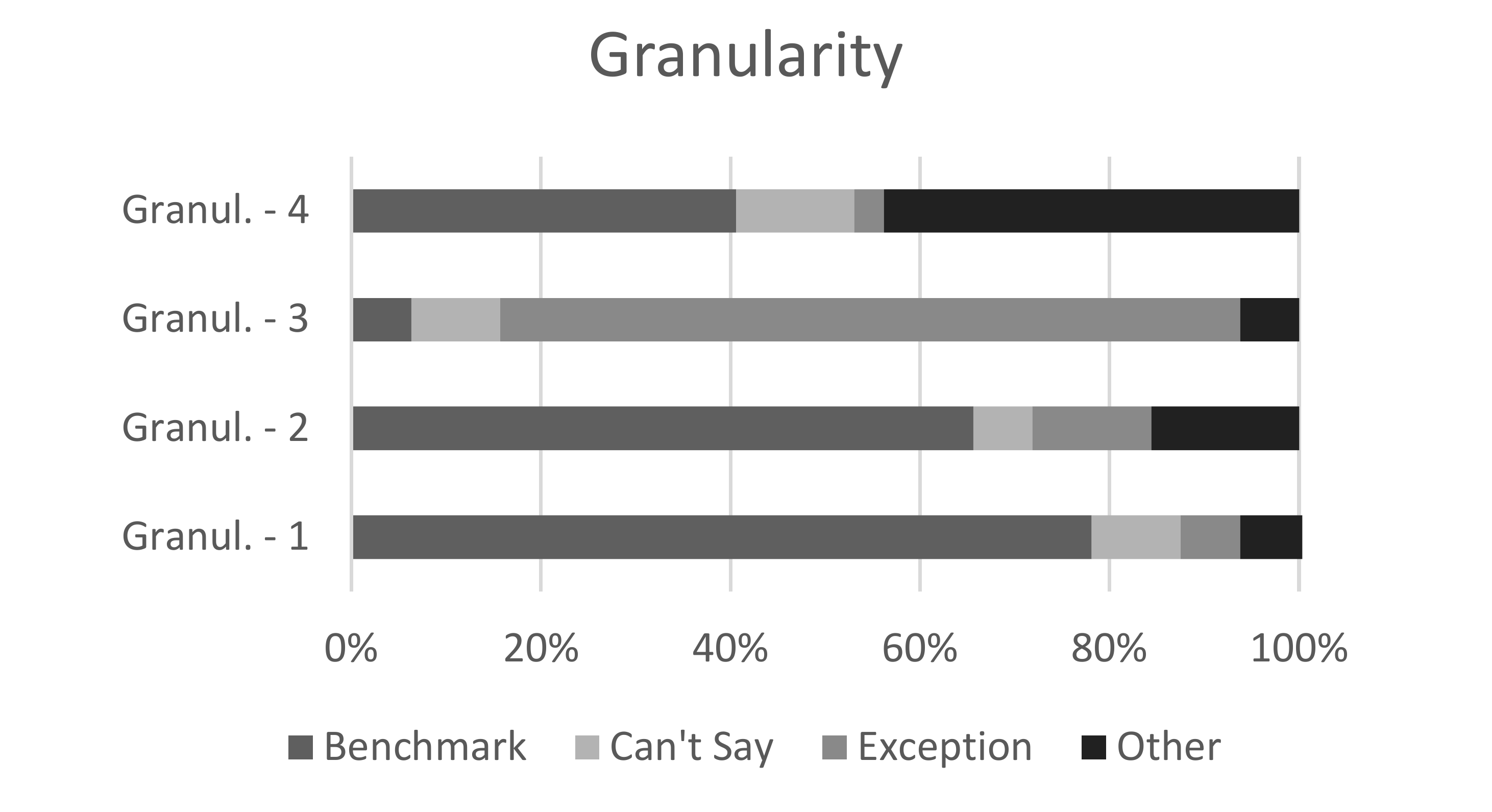}}
\caption{Results of the Experiment by Categories}
\vspace{-10mm}
\label{fig:results}
\end{figure}

The validation of our framework passes by the observation that, given a change in one of the dimensions of the framework, i.e. the variability area, there is a change in the outcome of the decision process. As we selected benchmark exercises that require basic default reasoning \cite{Lifschitz_1989}, and suggested that default reasoning is one adequate way to perceive how decision-making happens during elicitation of requirement, the following validation can be seen as the verification that our dimensions are relevant to consider during elicitation of requirements. In other words, observing significant differences in proportion of answers, for each dimension, results in the validation of dimensions for characterizing validity conditions of default rules used by stakeholders during elicitation. Table \ref{proportion} presents the proportions of answer for each tested levels of the framework's dimensions, and for each possible answer. Note that B stands for \textit{benchmark}, E for \textit{exception}, O for \textit{other} and A for \textit{abstention}. These proportions are also illustrated in Figure \ref{fig:results}.

Firstly, proportions of answers reported in Table \ref{proportion} show that subjects often achieve the highest proportion of benchmark, i.e. the correct answer, for the referent context. This was expected, as the referent level is the first benchmark problem on which subjects reason. From a general point of view, we observe that performance of subjects decreases when the characteristics of the objects used in the initial default rule change. As an illustration, consider the granularity dimension: we observe that subjects achieve nearly 80\% of benchmark for the referent question in which it is explained that the object is on average one hour late with important damages. When subjects are asked to reason about the same problem with same object but considering that it is similar to another object of the context, the performance drops to 10\%, and subjects prefer the exception answer.

\begin{table} [!ht]
\begin{center}
\begin{tabular}{|p{1.4cm}|c|c|c|c|p{1.7cm}|c|c|c|c|}
  \hline
   & B & E & O & A & & B & E & O & A\\
  \hline
  Item - 1 & .537 & .074 & .167 & .222 & Rule - 1 & .327 & .308 & .135 & .231\\
  Item - 2 & .370 & .204 & .111 & .315 & Rule - 2 & .385 & .250 & .115 & .250 \\
  Item - 3 & .278 & .148 & .296 & .278 & Rule - 3 & .269 & .288 & .019 & .423 \\
  Item - 4 & .352 & .130 & .278 & .241 & Rule - 4 & .173 & .500 & .154 & .173\\ \hline
  
  Loc. - 1 & .435 & .109 & .152 & .304 & Act. - 1 & .356 & .356 & .111 & .178\\
  Loc. - 2 & .326 & .239 & .087 & .348 & Act. - 2 & .400 & .333 & .156 & .111 \\
  Loc. - 3 & .239 & .261 & .239 & .261 & Act. - 3 & .400 & .333 & .156 & .111 \\
  Loc. - 4 & .065 & .348 & .130 & .457 & Act. - 4 & .133 & .422 & .067 & .378\\
 
  \hline
   Rel. - 1 & .500 &  .250 & .094 & .156 & Granul. - 1 & .788 & .091 & .061 & .091 \\
   Rel. - 2 & .375 &  .375 & .187 & .094 & Granul. - 2 & .667 & .061 & .121 & .152\\
  Rel. - 3 & .125 &  .313 & .406 & .156 & Granul. - 3 & .091 & .091 & .758 & .061 \\
  Rel. - 4 & .312 &  .500 & .031 & .156 & Granul. - 4 & .424 & .121 & .030 & .424 \\
  \hline
\end{tabular}
\caption{Proportion of Answers to our Questionnaires}\label{proportion}
\end{center}
\vspace{-10mm}
\end{table}

Secondly, it is interesting to note that in the case of activities and rules (the two dimensions for which the best performance is not achieved with the referent level), the highest proportion of benchmark answers is achieved for levels that are the most likely to happen for subjects. As an example, note that subjects reach the highest performance for level 3 of the activity dimension (trains were created for both people and freight transport), because they consider it more plausible than level 1 (trains were created only for freight). This observation was confirmed during a feedback session organized after the experiment.

Thirdly, we observe that subjects reach highest rate of abstention when the object that is described has characteristics that are impossible or difficult to reason about. This is for instance the case with level 2 of the Items dimension, in which the object is described as a teleportation means.

Significance tests presented in Table \ref{results} are performed using a repeated measure ANOVA on the previous proportions of answers, for each dimension and for each level. Working on the proportions implies that the answers selected by subject are considered as another potentially influencing factor \cite{Neter_1996}. In this experiment, we are therefore concerned with the interaction between answers and the levels of dimensions. Tests were performed with \textit{R-Project}, using the \textit{ezANOVA package}. Table \ref{results} summarizes the results for the \textit{answer} factor, and for the interaction between \textit{answers and levels} of the dimensions. The test is successful when we can reject the null hypothesis of the repeated measure ANOVA -- which states that $\mu_1 = \mu_2 = \mu_3 = \mu_4$ -- and accept the alternative hypothesis --according to which at least one mean is different to another mean. We observe a statistically significant effect of each dimensions level on the outcome of decision-making that requires default reasoning.

\begin{table} [!t]
\begin{center}
\begin{tabular}{|l|l|l|l|l|l|l|l|l|}
  \hline
  & \multicolumn{4}{|c|}{\textbf{Answer}} & \multicolumn{4}{|c|}{\textbf{Answer * Level}} \\ \hline
   \textbf{Dim.} & DF(n,d) & $\epsilon$ & F-Stat & P-Val & DF(n,d) & $\epsilon$ & F-Stat & P-Val \\ \hline
   Item & (3,159) & 1.00 & 8.510 & $0.000^{*}$ & (9,477) & 0.87 & 2.048 & $0.041^{*}$ \\
   Rule & (3,153) & 0.82 & 5.097 & $0.004^{*}$ & (9,459)  & 0.82 & 3.493 & $0.001^{*}$ \\ 
   Loc. & (3,135) & 0.89 & 5.547 & $0.002^{*}$ & (9,405) & 0.94 & 2.792 & $0.004^{*}$ \\
   Act. & (3,132) & 0.67 & 6.319  & $0.003^{*}$ & (9,396) & 0.87 & 2.756 & $0.006^{*}$ \\
   Rel. & (3,93) & 0.69 & 3.884 & $0.024^*$ & (9,279) & 0.92 & 3.793 & $0.000^{*}$ \\
   Gran. & (3,96) & 0.82 & 17.465  & $0.000^{*}$ & (9,79) & 0.76 & 17.381 & $0.000^{*}$ \\\hline
    \multicolumn{9}{|c|}{Significance codes:  `*' 0.05 `.' 0.1 ` ' 1}
     \\
  \hline
  \end{tabular}
\caption{Significance Test on Proportions of Answers}\label{results}
\end{center}
\vspace{-10mm}
\end{table}

\textbf{Normality assumption} was tested and verified for the set of data that we collected. Repeated measures ANOVA are however particularly susceptible to the violation of the \textbf{sphericity assumption}. This was the case for our data set. P-Values reported in Table \ref{results} are therefore the result of a Huynh-Feldt correction process \cite{Field_1998}. The correction amounts to multiplying freedom degrees by a parameter $\epsilon$, obtained via the sphercity test. This manipulation does not alter the value of the F-statistic, but results in an increase of the p-value, so that the risk of Type I error is reduced. Despite this correction, we observe a significant influence of each dimension on the outcome of a decision process requiring default reasoning. 

For each of the six questionnaires, we observe a significant influence of the ``answer'' factor. This is interpreted as the fact that answers B, E, O and A have been selected by subjects in significantly different proportions. Although interesting, this first observation does not validate our framework. Looking in the ``answers*levels'' columns, we observe that there is a significant influence. This can be interpreted as the fact that the proportion of an answer for B, E, O and A depends on the level of the dimension that is tested. This second observation validates our framework as a list of dimensions for characterizing validity conditions of default requirements.

\section{Discussion and Guidelines}
The objective of our experiment was to determine whether subjects change the way they use a default rule when the conditions initially defined for that default rule evolves according to one of the dimensions of our framework. The objective of the experiment was not to study the direction of these variations: we do not aim to explain why subjects choose the ``Benchmark'' answer for level 1 of a dimension and prefer the ``Exception'' answer for the second level. Such work, though interesting, would require many empirical efforts over the long term. 

We consider the objective of our experiment is achieved, because it enables to demonstrate the relevance of our framework as an adequate tool to consider context during the elicitation of requirements. This conclusion is achieved based on two major observations. Firstly, we observe that subjects systematically reach the maximum ``Benchmark'' proportion for the context that they consider to be the most representative of the real world. We interpret this as rationality, i.e., subjects are less likely to select a wrong answer for the benchmark problem when the conditions are close to their own perception of the real world. Secondly, we observe that subjects systematically decrease their performance when the context is changing according to one context dimension. We do not explain why they modify their choice, we simply observe that they significantly do so. We do not test the interactions between dimensions, we test the influence of standalone varying dimensions.

While the development of a methodology for the correct and systematic identification of default domain is beyond the scope of this paper, results of the experiment enable to suggest some relevant guideline to support engineers in such task:

\begin{itemize}
\item As decision-makers are more consistent when using default rules in situations that are conform to their perception of the world, we suggest that the elicitation of default requirements should take place in circumstances that are perceived as regular for the stakeholder. This implies discussing about well known objects, using well known elicitation techniques, avoiding references to unknown concepts or contexts;
\item Documentation should be considered as complete if and only if it describes in some way the conditions that must be verified for a default requirements, in terms of items that are targeted by the default requirement, but also localization, rules, relationships, activities and granularities of these items;
\item We observe subjects are more sensible to changes in some dimensions of the framework. Engineers should for example pay particular attention to documenting granularity, as it may lead to very important changes in the way default requirements is applied;
\item As the use of default rules is subjective, we suggest that engineers perform some cross-validation of identified default requirements with other stakeholders of the projects. This enables to reduce the margin for interpretation and confirm the correct documentation of the default domain.
\end{itemize}

\section{Conclusion}
In this paper, we study decision making during elicitation of requirements, and how to achieve better completeness of elicitation. Based on default logic, we propose a distinction between ground requirements that are explicitly communicated and default requirements that are implicitly assumed but not always correctly communicated to - or identified by - engineers. We argue the documentation of default requirements should account for the situation in which the default was designed, and claim such situation can be depicted using dimensions of context. We review definitions of context proposed in literature on context-aware and ubiquitous computing, and propose a context framework that supports engineers in the documentation of the default domain. We provide an empirical validation of the framework. We observe that each of the six categories of our framework significantly influences the way individuals make default based decisions. The framework therefore constitutes an adequate tool to support engineers in the identification and documentation of the default domain. We conclude the paper with a discussion about the results of the experiment, and some guidelines that can be suggested based on these results.

\bibliographystyle{splncs} 
\bibliography{bibfile1}

\end{document}